\documentclass{llncs}
\usepackage{graphicx,fancyhdr,amssymb,ifthen}
\usepackage[left=2cm,top=2cm,bottom=2cm,right=2cm,nohead,nofoot]{geometry}

\newcommand{\bR}{\mathbb{R}}

\newcommand{\bE}{\mathbb{E}}

\frontmatter

\begin{document}

\mainmatter              

\title{Conservation Law of Utility and Equilibria in Non-Zero Sum Games}
\author{Roman V. Belavkin\inst{1}}
\authorrunning{Roman Belavkin}   
%
\tocauthor{Roman Belavkin}
\institute{School of Engineering and Information Sciences\\
Middlesex University, London NW4 4BT, UK\\
Draft of 12 October 2010}

\maketitle              

\begin{abstract}
This short note demonstrates how one can define a transformation of a
non-zero sum game into a zero sum, so that the optimal mixed strategy
achieving equilibrium always exists.  The transformation is equivalent
to introduction of a passive player into a game (a player with a
singleton set of pure strategies), whose payoff depends on the actions
of the active players, and it is justified by the law of conservation
of utility in a game.  In a transformed game, each participant plays
against all other players, including the passive player.  The
advantage of this approach is that the transformed game is zero-sum
and has an equilibrium solution.  The optimal strategy and the value
of the new game, however, can be different from strategies that are
rational in the original game.  We demonstrate the principle using the
Prisoner's Dilemma example.\end{abstract}

\bibliographystyle{splncs}

Let $X$ and $Y$ be a dual pair of ordered linear spaces with respect
to bilinear form $\langle\cdot,\cdot\rangle:X\times Y\rightarrow\bR$.
If $x\in X$ is a utility function $x:\Omega\rightarrow\bR$, and $p\in
Y$, is a probability measure $p:\Omega\rightarrow[0,1]$, then the
expected utility is:
\[
\bE_p\{x\}=\langle x,p\rangle
\]
Consider a game such that the space of outcomes of the game is
$\Omega:=\Omega_1\times\cdots\times\Omega_m$, where $\Omega_i$ is the
set of pure strategies of $i$th player, and $x_i:\Omega\rightarrow\bR$
are the utility functions of the players.  In this case, the game is
zero-sum if and only if
\[
x_1+\cdots+x_m=0
\]
If $p_1,\dots,p_m\in Y$ are mixed strategies, then
\[
\sup_{p_1}\inf_{p_2,\dots,p_m}\bE_{p_1\times\cdots\times p_m}\{x_1\}
\leq\inf_{p_2,\dots,p_m}\sup_{p_1}\bE_{p_1\times\cdots\times p_2}\{x_1\}
\]
The famous Min-Max theorem \cite{Neumann-Morgenstern} states that in
zero-sum games, there always exists a mixed strategy $\bar
p_1\times\cdots\times\bar p_m$, called a {\em solution}, such that the
above holds with equality, and the common value is called the {\em
  value} of the game.

One of the problems in game theory is the existence of a solution to a
non-zero sum game.  Let us consider a game, such that
\[
x_1+\cdots+x_m\neq0
\]
If the utility functions add up to a constant function, so that $\sum
x_i\in\bR1:=\{\beta 1\in X:\beta\in\bR\}$, then one can add a constant
function $x_0:=-\frac1m\sum x_i$ to each $x_i$ so that the new
utilities $\tilde x_i=x_i+x_0$ are zero-sum.  We argue that the same
can be done in the case when $x_0$ is not a constant function.  Thus,
we define a bijection $T:X\rightarrow\tilde X$ by
\[
T(x):=x+x_0=x-\frac1m\sum_{i=1}^m x_i
\]
It is easy to see that the new utility functions are zero sum:
\[
\tilde x_1+\cdots+\tilde x_m=\sum_{i=1}^m(x_i+x_0)=\sum_{i=1}^mx_i+mx_0
=\sum_{i=1}^mx_i-\sum_{i=1}^mx_i=0
\]
Notice also that the new utility functions are proportional to the differences between the original utility functions and the sum of utilities of other players
\[
\tilde x_i=T(x_i)
=x_i-\frac1m\sum x_i
=\frac1m\left((m-1)x_i-\sum_{j\neq i}x_j\right)
\]
In zero-sum games, $\tilde x_i=x_i$ (i.e. $T(x)=x$), because $x_0=0$,
or equivalently the utility functions differ by some constant function
$x_i-x_i\in\bR1$.  Our transformation applies also to the case when
these differences are not constant functions.

The interpretation of adding utility $x_0$ to $x_i$ is as follows.  We
can extend $\Omega=\prod_{i=1}^m\Omega_i$ to
$\tilde\Omega:=\Omega_0\times\Omega$, where $\Omega_0:=\{0\}$ is a
singleton set so that $\tilde \Omega=\Omega_0\times\Omega=\Omega$
(because a singleton set plays the role of a unit with respect to
multiplication of sets).  The singleton set $\Omega_0=\{0\}$
represents a player with only one pure strategy, and we refer to it as
a {\em passive player}.  In another interpretation, $\Omega_0$ may
represent a player, whose pure strategy has already been chosen.  The
payoff $mx_0:\Omega\rightarrow\bR$ to this player, however, may be
non-constant, and depend on strategies of the other {\em active}
players (whose pure strategies have not yet been chosen).  We argue
that without including $\Omega_0$ and $x_0$ into the representation,
the total utility of active players $x_1+\cdots+x_m=-mx_0$ is not
constant, and therefore such a representation contradicts the {\em
  conservation law of utility} in a game:
\[
x_1+\cdots+x_m\in\bR1
\]
We argue that games, in which the above law is broken, such as the
non-zero sum games, have incomplete representation whereby some
passive player (or players) has not been taken into account.  The
complete representation is achieved by transformation $T(X)=X+x_0$,
and the transformed game is zero sum.  A solution to such a game
always exists, and its value is
\[
\bE_{\bar p_1\times\cdots\times\bar p_m}\{\tilde x_i\}=
\bE_{\bar p_1\times\cdots\times\bar p_m}\{x_i+x_0\}
\]
If $\bE_{\bar p_1\times\cdots\times\bar p_m}\{x_0\}=0$ (e.g. if
$x_0=0$), then this value equals the value of the original game.

\begin{example}[Prisonner's Dilemma]
Let us consider the classical example of two-person game, when
$\Omega=\Omega_1\times\Omega_2$, and each player (the prisoner) has
two pure strategies $\Omega_i=\{0,1\}$ --- cooperate ($\omega=0$) or
defect ($\omega=1$).  The payoff to each player is given by the
utility function $x_i:\Omega\rightarrow\bR$, which can be written in a
$2\times 2$-matrix form
\[
x_1=(x_{1ij})=\left(\begin{array}{rr}
-.6 & -10\\
0 & -5
\end{array}
\right)\,,\quad x_2=x_1^\dag
\]
The classical `rational' solution to this game is for each player to
use strategy $\omega=0$ (cooperate), so that $p_i(0)=1$, and
the value of the game is
\[
\bE_{0\times0}\{x_i\}=-.6
\]
However, when human participants play this game, they usually
cooperate or defect with almost equal probability so that the observed
strategies are dramatically different from $p_i(0)=1$.  Note that for
$p_i(0)=.5$, the expected payoff to each player becomes significantly
lower
\[
\bE_{.5\times.5}\{x_i\}=-3.9
\]
This `irrational' behaviour presents a paradox that defied many
attempts to explain it.

Observe that the game described is not zero-sum, because
\[
(x_{1ij})+(x_{2ij})=\left(\begin{array}{rr}
-1.2 & -10\\
-10 & -10
\end{array}\right)\quad\Rightarrow\quad
x_1+x_2\notin\bR1
\]
so that in fact the utilities $x_1$ and $x_2$ do not differ by any
constant function.  Let us now introduce the passive player
$\Omega_0=\{0\}$ (the detective), whose payoff is $2x_0=-(x_1+x_2)$,
and so $x_0$ is
\[
x_0=(x_{0ij})=-\frac12\Bigl[(x_{1ij})+(x_{2ij})\Bigr]=\left(
\begin{array}{rr}
.6&5\\
5&5
\end{array}\right)
\]
Thus, the payoff to the detective depends on the strategies of the
prisoners, and it is maximised if at least one of the prisoners
defects.  The utilities of the prisoners are transformed $T(x)=x+x_0$
to
\[
\tilde x_1=(\tilde x_{1ij})=\left(\begin{array}{rr}
0 & -5\\
5 & 0
\end{array}
\right)\,,\quad\tilde x_2=-\tilde x_1
\]
In this representation, the game becomes zero-sum, and it has solution
$\bar p_i(0)=.5$ and the value $\bE_{.5\times.5}\{\tilde
x_i\}=\bE_{.5\times.5}\{x_i+x_0\}=-3.9+3.9=0$.  This expected payoff
to each player is lower than that of the strategy $p_i(0)=1$, but it
is independent of the decision of the other player to defect.
\end{example}

\bibliography{rvb,nn,other,newbib,ica}

\begin{thebibliography}{1}

\bibitem{Neumann-Morgenstern}
von Neumann, J., Morgenstern, O.:
\newblock Theory of games and economic behavior. first edn.
\newblock Princeton University Press, Princeton, NJ (1944)

\end{thebibliography}

\end{document}